\documentclass[twocolumn,showpacs,aps,prl,superscriptaddress,floatfix]{revtex4}
\usepackage{graphics}
\usepackage{epsfig}
\usepackage{subfigure}

\newcommand{\be}{\begin{equation}}
\newcommand{\ee}{\end{equation}}
\usepackage{amsmath}

\usepackage{bm} 
\usepackage{amssymb}
\usepackage{natbib}
\usepackage{graphicx}
\usepackage{color}

\begin{document}

\title{Convection in multiphase flows using Lattice Boltzmann methods}
\author{L. Biferale} \affiliation{Dept. of Physics and INFN, U. of Tor
  Vergata Via della Ricerca Scientifica 1, 00133 Rome, Italy}
\author{P. Perlekar} \affiliation{Department of Physics and Department
  of Mathematics and Computer Science, Eindhoven University of
  Technology, 5600 MB Eindhoven, the Netherlands}
\author{M. Sbragaglia} \affiliation{Dept. of Physics and INFN, U. of
  Tor Vergata Via della Ricerca Scientifica 1, 00133 Rome, Italy}
\author{F. Toschi} \affiliation{Department of Physics and Department
  of Mathematics and Computer Science, Eindhoven University of
  Technology, 5600 MB Eindhoven, the Netherlands and\\IAC, CNR, Via dei
  Taurini 19, 00185, Roma, Italy and\\INFN, via Saragat 1, I-44100
  Ferrara, Italy.
}

\begin{abstract}
  We present high resolution numerical simulations of convection in
  multiphase flows (boiling) using a novel algorithm based on a
  Lattice Boltzmann method. We first validate the thermodynamical and
  kinematical properties of the algorithm. Then, we perform a series
  of 3d numerical simulations at changing the mean properties in the
  phase diagram and compare convection with and without phase
  coexistence at $Ra \sim 10^{7}$. We show that in presence of
  nucleating bubbles non-Oberbeck Boussinesq effects develops, mean
  temperature profile becomes asymmetric, heat-transfer and
  heat-transfer fluctuations are enhanced. We also show that
  small-scale properties of velocity and temperature fields are
  strongly affected by the presence of buoyant bubble leading to high
  non-Gaussian profiles in the bulk.
\end{abstract}

\maketitle

Thermal convection, the state of a fluid heated from below and cooled
from above, is an ubiquitous phenomena in nature, present in many
industrial and geophysical applications both at micro and macro-scales
\cite{some_general_paper_geophysical}. It is also challenging from the
theoretical point of view, due to its extremely reach and different
regimes ranging from intricate pattern formations at small temperature
difference between bottom and top plates (i.e. moderate Rayleigh
number) to extremely turbulent behavior where heat transfer and its
adimensional definition (i.e. Nusselt number) is dominated by bulk or
boundary layer physics (or by both, see e.g. recent reviews
\cite{bodenschatz_arfm}).  Most of the time thermal convection is
studied in its simplest version, the so-called Oberbeck-Boussinesq
(OB) approximation, where a single phase --unstratified-- fluid is
present with constant material properties. Compressibility is also
neglected except for buoyancy forces. Needless to say, in many
situation some, or all, of the above assumption breaks down and one
enters in the realm of Non-Oberbeck-Boussinesq (NOB)
convection. Deviations from OB can arises in many different way. Two
notable cases are (i) the presence of stratification, as in many
geophysical applications (ii) and under boiling conditions, i.e. when
the parameters excursion inside the convective cell allows for phase
coexistence \cite{boiling_ahlers,boiling_general}.
\begin{figure}[t]
\begin{center}
  \includegraphics[scale=.7]{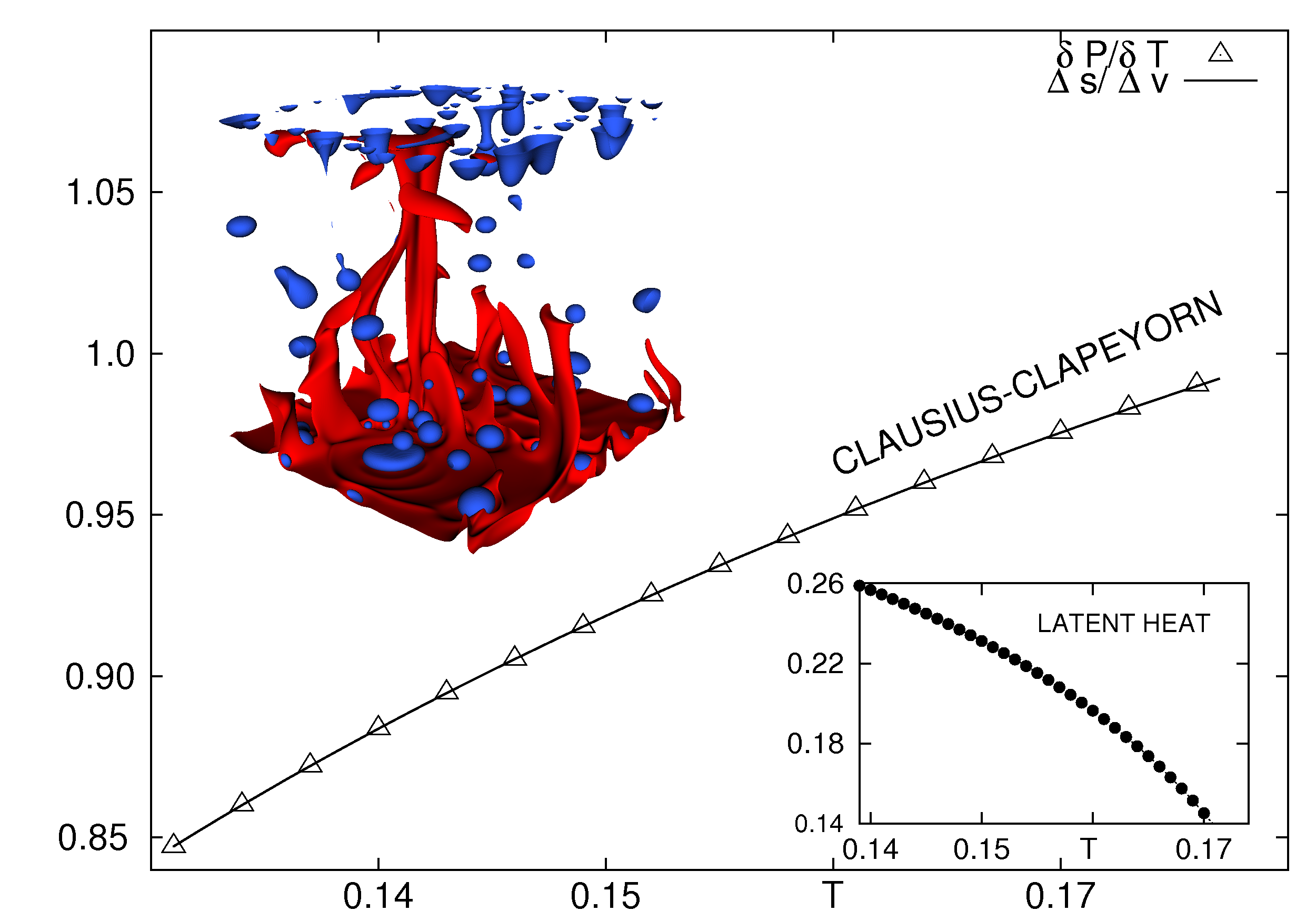}
\end{center}
\caption{Check of Clausius-Clapeyron relation $\partial P/\partial T = {\Delta s}/{\Delta v}$ in our numerical scheme at varying $T/T_c$. $P$ is the equilibrium pressure at coexistence temperature $T$, $v=1/\rho$ is the specific volume, $T_c$ is the critical temperature and $s(T,\rho)$ is the specific entropy. Bottom inset: latent heat,  $\lambda=T \Delta s$ vs $T$. Top inset: Bubbles are in blue. Regions with high temperature are in red. The system has no-slip velocity at bottom and top walls and it is periodic on the horizontal directions. 
  \label{fig:1}}
\end{figure}

In this paper we address the latter case. We study thermal convection
in a 3d cell in a high turbulent regime where large bubbles (larger
than the turbulent viscous scale) can nucleate in the layer close to
the bottom wall with a non-negligible heat-exchange between liquid and
vapor. To do that, we present, validate and apply a novel numerical
scheme based on a diffuse interface Lattice Boltzmann method (LBM)
\cite{succi,zhang_chen}. We are not restricted to treat bubbles as
point-like \cite{verzicco} and we fully resolve the thermo
hydrodynamical properties of the gas and liquid phases. Beside the
methodological aspects, we also address physical questions connected
to the enhancement/depletion of heat flux in presence of bubbles,
statistics of mean global properties as well as small-scales effects
for both velocity and temperature fluctuations. We present two series
of high-resolution numerical simulations up to $512^3$ collocation
points at $Ra\sim 10^{7}$ with and without phase coexistence, such as
to be able to directly compare on the same geometrical set up the
effect of boiling on convection.
\begin{figure}
\begin{center}
\includegraphics[scale=.7]{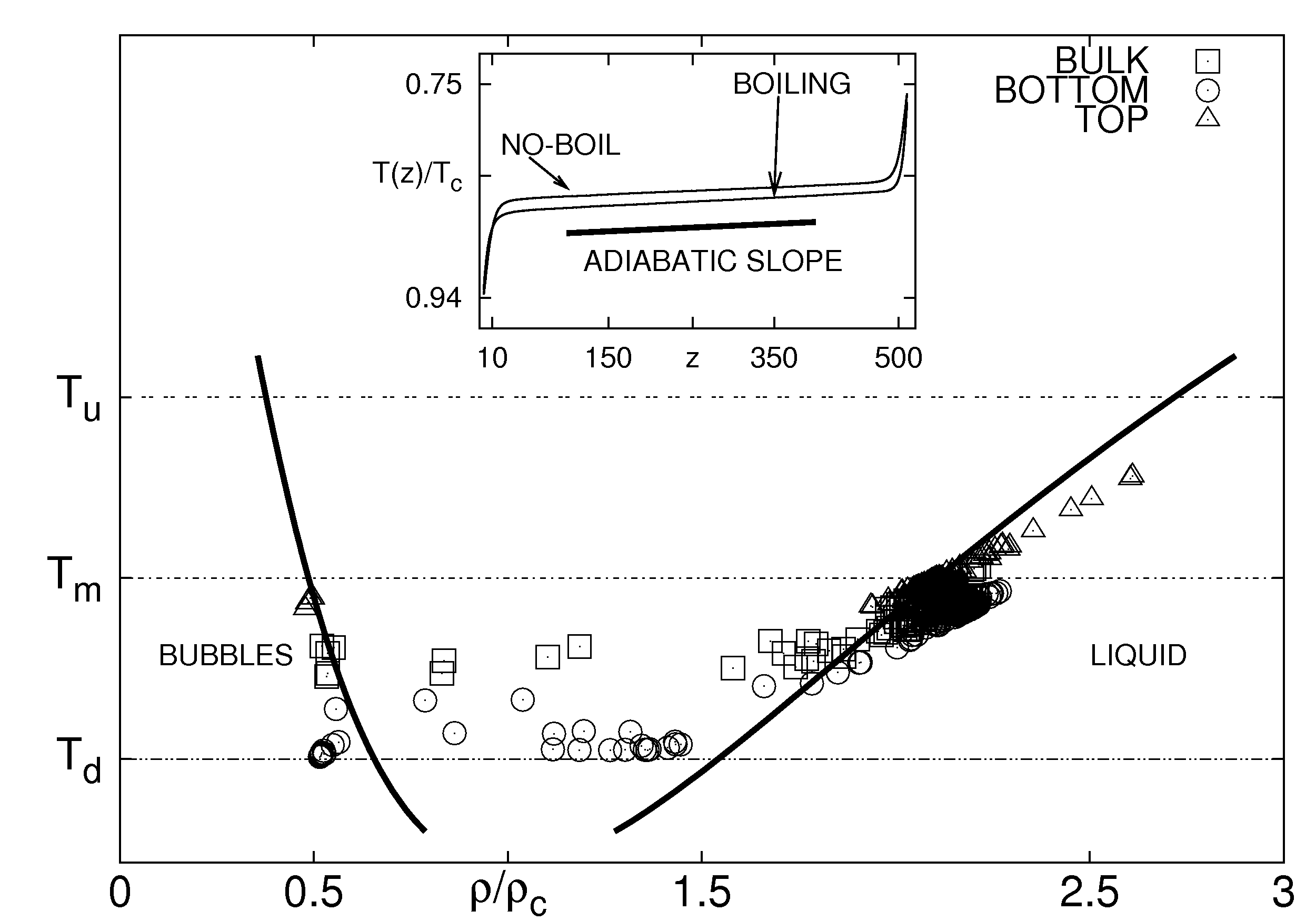}
\includegraphics[scale=.7]{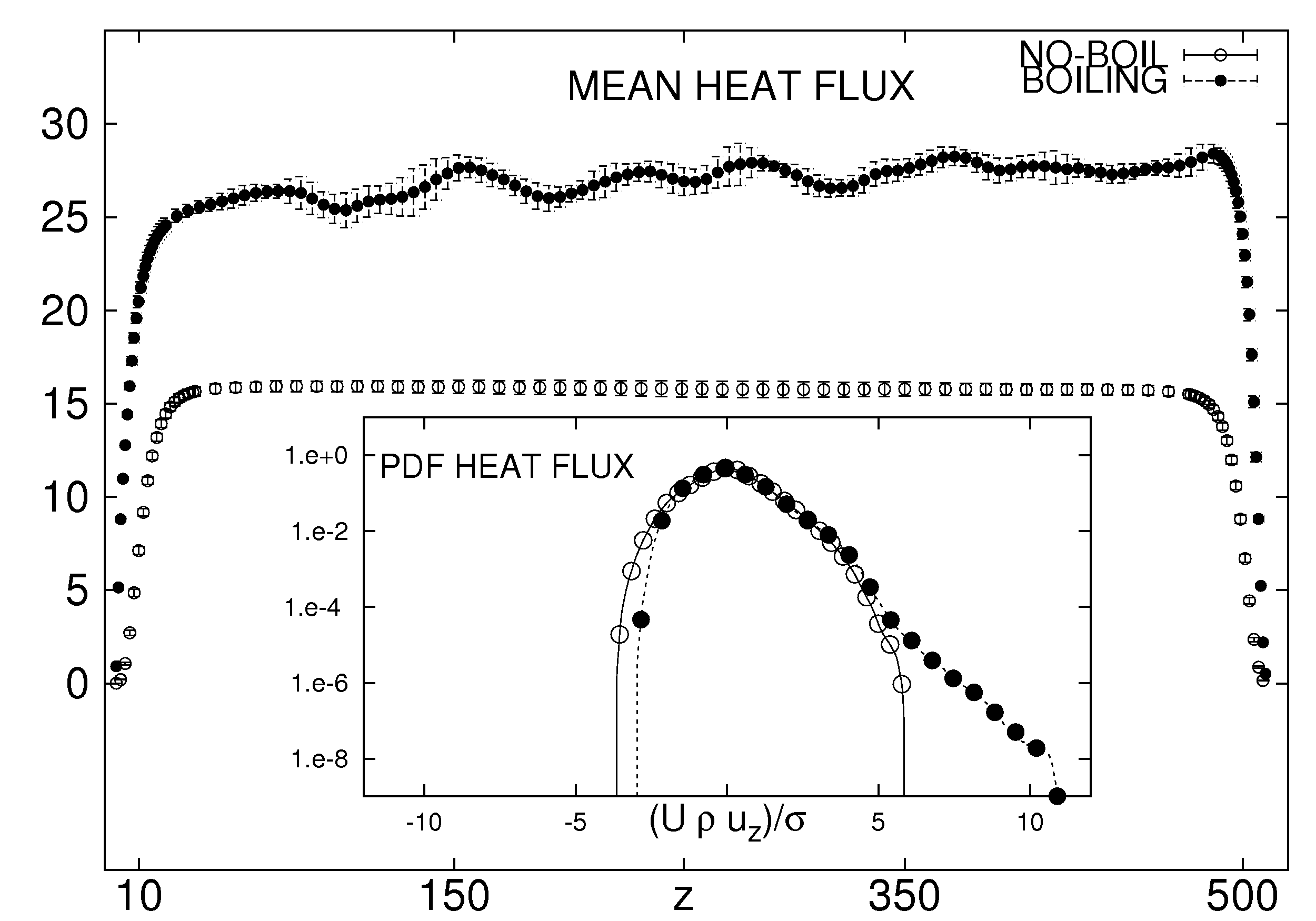}
\end{center}
\caption{{\sc Top panel.} 
Phase space $T-\rho$ equilibrium curves (solid lines) 
superposed with a scatter plot of temperature and density values measured  under boiling condition (both made dimensionless using the critical values, $T_c$ and $\rho_c$).
Notice the presence of bubbles with different temperature inside the volume. Different symbols corresponds to measurements taken in top, bottom or bulk region. The horizontal dashed line correspond to top, $T_u$, bottom, $T_d$, and mean, $T_m$, temperatures. Inset: 
 mean temperature profile, $\overline{T(z)}/T_c$ vs $z$ (in lattice units) for boiling and non-boiling conditions. The straight line corresponds to the adiabatic slope.\\
{\sc Bottom panel.} Bulk contribution to the heat flux normalized to its diffusive value, $\overline{ \rho U u_z}/[\kappa (T_d-T_u)/L]$, (Nusselt) for boiling and non-boiling case at comparable Rayleigh number. 
Inset: pdf of $\rho U u_z$ normalized to have mean area and mean variance for both boiling and non-boiling cases. 
 \label{fig:2}
}
\end{figure}
With respect to experimental studies, numerical simulations offer the
unique advantages to allow access to all quantities without affecting
the fluid dynamics and to confine the fluid inside ``ideal'' surfaces
(i.e. perfect thermal properties at the wall, perfect smoothness of
the boundaries etc...)  On the other hand, a limitation consists in
the difficulty to reach high Rayleigh numbers and to push the physical
parameters as density contrast, interface thickness, viscosity and
thermal diffusivity to realistic situations. This is why the interplay
between numerical investigations and experiments is crucial in this
field.\\
The equations of motion describing a non-ideal fluid in presence of
thermal fluctuations are:
\begin{equation}
  \partial_t \rho u_i  +  \partial_j(\rho u_i u_j) =
  -\partial_i P +  \partial_j (\mu  (\partial_i u_j +\partial_j u_i)) + g \rho \hat z
\label{eq:ns}
\end{equation}
where $\mu = \rho \nu$ is the molecular viscosity, $g$ is the gravity,
$\rho$ is the local fluid density and $P(\rho,T) = P_0(\rho,T) +
P_{NI}(\rho)$ is the non-ideal pressure. Pressure is fixed by the
equation of state and it is made of two terms, the ideal part
$P_0(\rho,T) = \rho T$ and the non ideal part which in our LBM system
reads: $P_{NI}(\rho) = G \exp(-2/\rho)$ (see below).  The equation for
the internal energy, $U = c_v T + \int d \rho P_{NI}/\rho^2 $ is given
by: \be
\label{eq:internal}
\rho D_t U + P \partial_j u_j = \kappa \partial_{jj} T \ee where
$\kappa$ is the thermal conductivity. The above equation can also be
rewritten as one of the two following equivalent expressions:
\begin{equation}
\begin{cases}
\label{eq:eq}
c_p \rho D_t T -\beta T D_t P = \kappa \partial_{jj} T \\
c_v \rho D_t T + P_0 \partial_j u_j = \kappa \partial_{jj} T
\end{cases}
\end{equation}
where $D_t$ stands for the material derivative, $c_v$ is the specific
heat at constant volume, $c_p$ and $\beta = - (\partial_T \rho)/\rho$
are the specific heat and compressibility at constant pressure,
respectively.  The above set of equations tends to the usual OB system
when the fluid is considered single phase, incompressible and both
$\mu, \kappa$ are constant \cite{spiegel-veronis}. In table 1 we
report the characteristic values for all relevant parameters, with and
without boiling.

Because of bubble nucleation/evaporation, a key role is played by the
$D_tP$ term in (\ref{eq:eq}). Take for example a convective cell of
eight $L$ with imposed temperature, $T_d$, at the bottom wall and
$T_u$ at the top wall. Then, the heat balance across an horizontal
layer at distance $z$ from the bottom wall is given by the expression:
\begin{equation}
  \label{heat2}
  {\partial_t \overline{\rho U}|_z +  \partial_z \overline{\rho U v_z  - \kappa \partial_{z} T}}|_z = - \overline{P\partial_j u_j}|_z
\end{equation}
where with $\overline{(\cdot)}|_z$ we intend a spatial average at
fixed $z$. In a stationary situation, we can define a $z$-dependent
dimensional Nusselt number $ Nu(z) = \overline{\rho U v_z -
  \kappa \partial_{z} T}|_z $ which satisfy an integral constraint \be
Nu(z) -Nu(0) = -\int_0^z dz' \overline{P \partial_j u_j}|_{z'}.
\label{eq:flux} \ee With the above definition, the Nusselt number is
not anymore constant throughout the cell, we may exchange heat by
nucleating and evaporating bubbles or by simple compressible effects
inside each phase.

\begin{table}
\begin{center}
\begin{tabular}{|c | c c c c c c c c|}
  \hline    
  & $\Delta \tilde T$ & $\Delta \tilde c_p$ & $\chi_g$ & $\chi_l$ & $\nu$    & $Ra$           & $\Delta \tilde \beta$ & $Pr$ \\
  \hline 
  BOILING  & $0.226$             & $1.2$            & $0.008$ & $0.0018$ & $0.0165 $& $3 \times 10^{7}$ & $1.6$                     & $9$ \\
  NO-BOIL. & $0.230$             & $0.2$            &   $--$       & $0.0018$ & $0.0165$ & $2 \times 10^{7}$ & $0.1$                     & $9$ \\ 
  \hline
\end{tabular}
\caption{$\Delta \tilde T $, $\Delta \tilde c_p$, $\Delta \tilde \beta$: values of temperature, liquid  heat capacity and liquid compressibility difference between the two walls (all normalized with their respective values at the center of the cell, $z_c$). $\chi_{l,g}=\kappa/(c_p(z_c) \rho_{l,g}(z_c))$: 
thermal diffusivity of liquid (l) and gas (g). $\nu$: kinematic viscosity. Rayleigh  $Ra$ and Prandtl $Pr$ numbers are evaluated at $z_c$ and in the liquid phase:
$ Ra  = \frac{g \beta(z_c) L^4 (\Delta T/L - \gamma_{ad}(z_c))}{k/(\rho(z_c)c_P(z_c)) \nu }$ and  $Pr = \frac{\nu}{\chi^{(l)}(z_c)}$
 where $L=512$ is the  cell height (in grid units) and $\gamma_{ad} = \beta T g/c_p$ is the adiabatic gradient.  The Jacob number quantifying the ration between the sensible heat and the latent heat \cite{verzicco} is $Ja \sim 3$.  
 \label{table:param} }
\end{center}
\end{table}

{\sc Algorithm}. The numerical algorithm used is based on discrete
kinetic models \cite{succi}. The starting point is a standard coupled
mesoscopic dynamics described by \cite{succi,Guo}:
\begin{eqnarray}
\label{LBEQ}
f_{l}({\bm x}+{\bm c}_l,t+1)-f_{l}({\bm x},t)=&-\frac{1}{\tau_\nu} (f_l-f^{(eq)}_l )({\bm x},t) \\
g_{l}({\bm x}+{\bm c}_l,t+1)-g_{l}({\bm x},t)=&-\frac{1}{\tau_\kappa} (g_l-g^{(eq)}_l)({\bm x},t)
\label{LBEQ2}
\end{eqnarray}
where $f_{l}({\bm u},t)$, $g_{l}({\bm x},t)$ stand for the probability
density functions to find at $({\bm x},t)$ a particle whose kinetic
velocity belongs to a discrete and limited set ${\bm c}_l$ (with
$l=1,19$ in the $D3Q19$ LBM adopted here \cite{succi}).  Density,
momentum and temperature are defined as coarse-grained (in velocity
space) fields of the distribution functions \be\label{HYDRO}
\rho=\sum_{l}f_{l} \hspace{.2in} \rho {\bm u}=\sum_{l}{\bm c}_l f_{l}
\hspace{.2in} \rho T=\sum_{l} g_{l}.  \ee The local kinetic equilibria
$f^{(eq)}_l({\bm u^{\prime}},\rho)$ and $g^{(eq)}_l({\bm u},{\bm F},
T)$ are usually expanded in suitable polynomial basis \cite{SC93} in
such a way that a Chapman-Enskog expansion \cite{Guo} leads to the
equations for density, momentum and temperature
(\ref{eq:ns})-(\ref{eq:eq}): the streaming step on the left hand side
of (\ref{LBEQ}) reproduces the inertial terms in the hydrodynamical
equations, whereas dissipation and thermal diffusion are connected to
the relaxation (towards equilibrium) properties in the right hand
side, with $\nu$ and $\kappa$ related to the relaxation times
$\tau_\nu$, $\tau_\kappa$ \cite{succi}.  Non ideal thermodynamics is
obtained by a well controlled procedure shifting the velocity in the
equilibrium distribution, ${\bm u^{\prime}}={\bm u}+\tau_\nu {\bm
  F}/\rho$, with a forcing term mimicking the effect of an internal
pseudo-potential \cite{SC93,Guo}. In particular, we adopt the standard
form: \be\label{FORCE} {\bm F}=-{\cal G}\sum_{l=1}^{N} w(|{\bm
  c}_l|^2){\bm c}_l \psi[\rho({\bm x})]\psi[\rho({\bm x}+{\bm c}_l)]
\ee where ${\cal G}$ is a parameter dictating the overall strength of
the non-ideal interactions.  The weights $w(|{\bm c}_l|^2)$ are used
to enforce isotropy up to the 4th order in the velocity tensors
\cite{pre}. The pseudo-potential, $\psi[\rho]$, encompasses the
macroscopic effects of both long-range attraction and short-range
repulsion.  Although various choices have been presented for the
choice of $\psi[\rho]$ \cite{Yuan,Kupershtokh}, here it is crucial to
set it to $ \psi[\rho]=\exp(-1/\rho)$, in such a way to reproduce the
thermodynamic consistency on the lattice \cite{SBRAGAGLIASHAN} (see
fig. \ref{fig:1}).  Furthermore, to reproduce the correct ideal part
of the pressure, $P_0=\rho T$, a coupling between $f_l$ and $g_l$
populations in (\ref{eq:eq}) is needed. To this end, we used a recent
proposal \cite{ZhangTian08} by plugging the dynamical temperature $T$
extracted from the population $g_l$ in the equilibrium distribution of
(\ref{LBEQ2}): in the limit of vanishing interaction, this is
equivalent to impose a second order momentum of $f^{(eq)}_l$ equal to
$\sum_{l} f^{(eq)}_l c^i_l c^j_l =\rho T \delta_{ij}+\rho u_i
u_j$. Finally, in order to reproduce exactly the divergence term,
$P_0 \partial_j u_j$, in (\ref{eq:eq}), we found necessary to add a
proper counter term to the evolution of $g_l$ populations in
(\ref{LBEQ2}), as proposed in \cite{PRASIANAKIS}.  As a result, we
ended with a LBM scheme able to reproduce in the hydrodynamical limit
the NS equations (\ref{eq:ns}-\ref{eq:eq}) with a non-ideal Pressure
tensor and a consistent definition of latent heat (see
fig. {\ref{fig:1}).

  {\sc Single point quantities.}  In fig. (\ref{fig:2}) we shown a
  scatter plot of $T({\bf x},t)$ vs $\rho({\bf x},t)$ for a boiling
  cell. As one can see most of the volume is at {\it thermodynamical
    equilibrium}, superposing with the equilibrium curves in the
  $T-\rho$ phase space. The presence of bubbles is clearly detected by
  the spots concentrating along the vapor branch and it is also
  interesting to notice that the corresponding bubble temperature is
  always larger than the mean temperature in the cell, indicating that
  bubbles are transferring temperature upwards very efficiently.
  Moreover, the temperature profile across the cell,
  $\overline{T(z)}$, becomes slightly asymmetric in presence of
  bubbles, a phenomenon also observed in other liquid-like NOB systems
  \cite{nob-temperature-profile}. Breaking of the top-down symmetry
  must not surprise. In particular, $\beta$ is not constant across the
  cell (i.e. density and temperature fluctuations are not strictly
  proportional as in OB) and $c_P$ decreases going from bottom to
  top. Both effects may have an impact on the averaged profiles as
  discussed and observed also in \cite{nob-temperature-profile}.
  Here, the temperature mismatch between the values at the center and
  the mean temperature is $1\%$ (inset of top panel in
  fig. \ref{fig:2}).
  Notice also, in the same plot, that  $\overline{T(z)}$ agrees with the expected profile given by the adiabatic gradient, due to the presence of a small stratification.  \\

  In the bulk, the heat flux (\ref{eq:flux}) is dominated by the
  convective term $\overline{\rho U u_z}$.  In bottom panel of
  fig. (\ref{fig:2}) we compare the Nusselt number for the boiling and
  non-boiling cell at comparable Rayleigh.  Two effects show
  up. First, heat flux is enhanced. Second, fluctuations around its
  mean profile are larger in presence of bubbles. We interpret this as
  a clear signature of the importance of the bubble dynamics in
  transporting heat between the two walls. This is the combined effect
  of temperature entrainment inside bubbles leveraged with the
  buoyancy effect that drift bubble upward much more efficiently then
  for plumes in single-phase convection. Because bubbles are rare in
  our system, this also implies an increase in heat flux fluctuations,
  as can be seen in the inset of bottom panel in fig. (\ref{fig:2})
  where we show the probability density function (pdf) of the heat
  flux measured only in the bulk cell. Clearly, the right tail are
  enhanced, due to bubble buoyancy. An open question is to see whether
  this enhancement of heat-flux is robust at changing Rayleigh and/or
  position in the phase diagram (with more or less bubble nucleation
  in the system). This needs to be investigated with new computational
  effort and will be reported in the future.

  {\sc Small-scales properties.} Buoyant bubbles brings information
  from the physics of the bottom boundary layer in the bulk of the
  system. We then expect also in the bulk an increase of small scales
  fluctuations concerning both velocity and temperature fields.  In
  fig. (\ref{fig:3}) we show the structure functions for vertical
  velocity and temperature:
\begin{equation}
  \begin{cases}
    S^{(p)}_{u_z}(r) = \langle [u_z({\bf x}+ {\bf r})-u_z({\bf x})\cdot \hat {\bf r}]^p \rangle_{bulk}\\
    S^{(p)}_{T}(r) = \langle [T({\bf x}+ {\bf r})-T({\bf x})\cdot \hat
    {\bf r}]^p \rangle_{bulk}
  \end{cases}
\end{equation}
where the average is restricted on all points ${\bf x}$ in the bulk of
the cell and the increment ${\bf r}$ is always taken in horizontal
directions. In the two panels of fig.(\ref{fig:3}) we show the results
for both quantities for $p=2$.  For both fields we have a viscous
range very well resolved, where the structure functions goes as
$\propto r^2$. Therefore, the presence of large-bubbles do not destroy
the differentiability at small scales, another signature that the
numerical set-up is under control. Second, boiling system have an
enhanced signal at small scales, meaning that energy dissipation is
globally increased. Third, for the boiling case we start to see an
inertial range with K41-like scaling $\propto r^{2/3}$. In the inset
of both panels we measure the Flatness (or Kurtosis) of each field
$K_{u_z,T} (r) = S_{u_z,T}^{(4)}(r)/ [S_{u_z,T}^{(2)}(r)]^2$ at
different scales, e.g. a way to quantify how much the pdf is
close/different to a Gaussian. Intermittency, as measured by the
deviation of the flatness from its Gaussian value, $K=3$ becomes more
and more important at decreasing the scale, in agreement with the
general observation that bubbles induce an increase of fluctuations in
the system. For temperature (top panel fig. \ref{fig:3}) the inertial
range behavior is much more singular than the case for velocity, due
to the enhancement of temperature jumps between inside and outside
bubbles. At difference from the case for velocity, where large scale
pdf is indistinguishable from a Gaussian ($K_{u_z} \sim 3$ for $r \sim
L$), here temperature is more sensitive to the presence of bubbles
also at large scale.

\begin{figure}[t]
\begin{center}
\includegraphics[scale=.7]{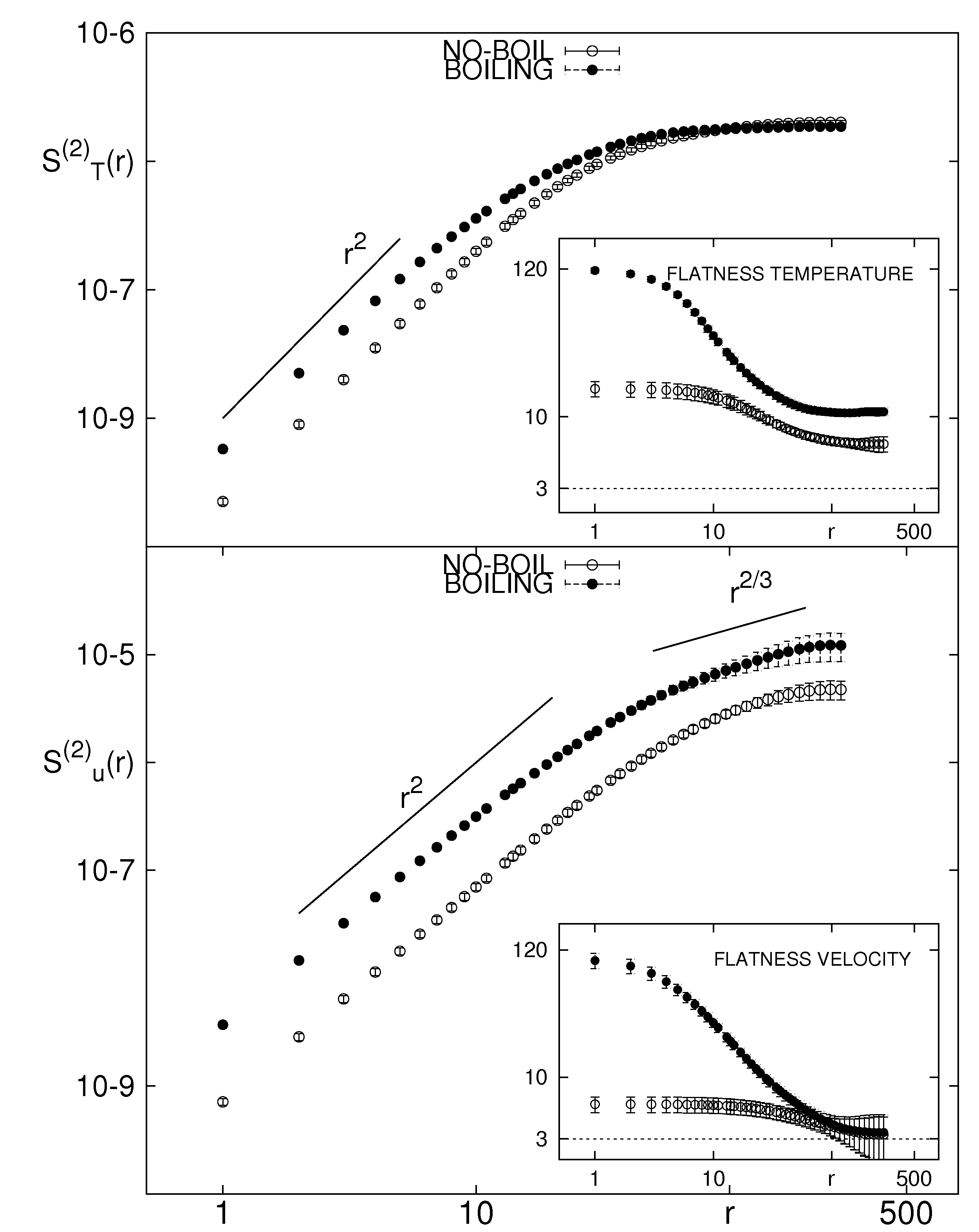}
\end{center}
\caption{2nd order Structure Function for velocity (bottom) and
  temperature (top) vs $r$ (in lattice units) for boiling and
  non-boiling systems. Inset: values of Flatness (same
  symbols)) \label{fig:3}}
\end{figure}

In conclusion, we have proposed and validated a novel LBM to attack
multiphase flows with full consistent definition of heat exchange in
the system (latent heat).  We have applied this scheme to study
convection under boiling condition and we have studied the effects of
nucleating large bubbles at the bottom boundary layer on both single
point observable (temperature profile and heat-flux) and two-point
correlation functions (structure functions). The latter, allowed us to
assess also the importance of bubbles on small-scales velocity and
temperature fluctuations, indicating an enhancement of the deviations
from Gaussian statistics with respect to the non-boiling case. We
acknowledge useful discussion with R. Verzicco.  The COST Action
MP0806 is kindly acknowledged.  Support within the DEISA Extreme
Computing Initiative is acknowledged (FP6 project RI-031513 and FP7
RI-222919).

\end{document}